\documentclass[aps,showpacs,preprint,superscriptaddress,endfloats*]{revtex4-1}
\usepackage{graphicx}
\usepackage{dcolumn}
\usepackage{bm}
\usepackage{amssymb}
\usepackage{amsmath}
\usepackage{epstopdf}
\usepackage{datetime}
\usepackage{makecell}
\usepackage[colorlinks,plainpages=false,linkcolor=blue,urlcolor=blue,citecolor=blue,pdfpagemode=UseNone,pdfstartview=FitBH]{hyperref}
\usepackage{charter}
\usepackage{hyperref}
\usepackage{bookmark}
\usepackage{subfigure}
\usepackage{multirow}
\usepackage{float}

\newcommand{\hmn}[1]{
  \ensuremath{\begingroup\setupHMN #1\endgroup}%
}

\newcommand{\setupHMN}{%
  \doHMN{-}{\HMNoverline}%
  \doHMN{*}{\HMNminverse}%
  \doHMN{i}{\infty}
}

\newcommand{\doHMN}[2]{%
  \begingroup\lccode`~=`#1
  \lowercase{\endgroup\let~}#2%
  \mathcode`#1="8000
}

\newcommand{\HMNminverse}[1]{\frac{#1}{m}}
\newcommand{\HMNoverline}[1]{\mkern1mu\overline{\mkern-1mu#1\mkern-1mu}\mkern1mu}

\begin{document}

\title{Revisiting the Magnetic Structure and Charge Ordering in La$_{1/3}$Sr$_{2/3}$FeO$_3$ by Neutron Powder Diffraction and M\"ossbauer Spectroscopy}
\author{F.~Li}
\email[]{Fei.Li@psi.ch}
\affiliation{Laboratory for Multiscale Materials Experiments, Paul Scherrer Institut, 5232 Villigen, Switzerland}
\author{V.~Pomjakushin}
\affiliation{Laboratory for Neutron Scattering and Imaging, Paul Scherrer Institut, 5232 Villigen, Switzerland}
\author{T.~Mazet}
\affiliation{Institut Jean Lamour, UMR-CNRS 7198, Universit\'e de Lorraine, BP 70239, 54506 Vandoeuvre-l\`es-Nancy Cedex, France}
\author{R.~Sibille}
\affiliation{Laboratory for Neutron Scattering and Imaging, Paul Scherrer Institut, 5232 Villigen, Switzerland}
\author{B.~Malaman}
\affiliation{Institut Jean Lamour, UMR-CNRS 7198, Universit\'e de Lorraine, BP 70239, 54506 Vandoeuvre-l\`es-Nancy Cedex, France}
\author{R.~Yadav}
\affiliation{Laboratory for Scientific Developments and Novel Materials, Paul Scherrer Institut, 5232 Villigen, Switzerland}
\author{L.~Keller}
\affiliation{Laboratory for Neutron Scattering and Imaging, Paul Scherrer Institut, 5232 Villigen, Switzerland}
\author{M.~Medarde}
\affiliation{Laboratory for Multiscale Materials Experiments, Paul Scherrer Institut, 5232 Villigen, Switzerland}
\author{K.~Conder}
\affiliation{Laboratory for Multiscale Materials Experiments, Paul Scherrer Institut, 5232 Villigen, Switzerland}
\author{E.~Pomjakushina}
\affiliation{Laboratory for Multiscale Materials Experiments, Paul Scherrer Institut, 5232 Villigen, Switzerland}

\date{\today}

\begin{abstract}

The magnetic ordering of La$_{1/3}$Sr$_{2/3}$FeO$_3$ perovskite has been studied by neutron powder diffraction and $^{57}$Fe M\"ossbauer spectroscopy down to 2 K. From symmetry analysis, a chiral helical model and a collinear model are proposed to describe the magnetic structure. Both are commensurate, with propagation vector \textbf{k} = (0,0,1) in \hmn{R-3c} space group. In the former model, the magnetic moments of Fe adopt the magnetic space group \hmn{P3_{2}21} and have helical and antiferromagnetic ordering  propagating along the \textit{c} axis. The model allows only single Fe site, with a magnetic moment of 3.46(2) $\mu_{\rm{B}}$ at 2 K. In the latter model, the magnetic moments of iron ions adopt the magnetic space group \hmn{C2/c} or \hmn{C2'/c'} and are aligned collinearly. The model allows the presence of two inequivalent Fe sites with magnetic moments of amplitude 3.26(3) $\mu_{\rm{B}}$ and 3.67(2) $\mu_{\rm{B}}$, respectively. The neutron diffraction pattern is equally well fitted by either model. The M\"ossbauer spectroscopy study suggests a single charge state Fe$^{3.66+}$ above the magnetic transition and a charge disproportionation into Fe$^{(3.66-\zeta)+}$ and Fe$^{(3.66+2\zeta)+}$ below the magnetic transition. The compatibility of the magnetic structure models with the M\"ossbauer spectroscopy results is discussed.

\end{abstract}


\maketitle

\section{INTRODUCTION}

The R$_{1/3}$Sr$_{2/3}$FeO$_3$ (R = rare earth) family is reported to show a crossover between localized and itinerant behavior by variation of the size of the rare earth ion \cite{Park1999}. For R = La, Pr and Nd, a 2Fe$^{4+}$ $\rightarrow$ Fe$^{3+}$ + Fe$^{5+}$ charge disproportionation (CD) accompanied by Fe$^{3+}$/Fe$^{5+}$ charge ordering (CO), a magnetic ordering, and a metal-insulator (MI) transition was reported to occur at 200, 180 and 165 K, respectively. For smaller rare earth ions no MI transition is observed, the compounds being purely insulating below room temperature.

\begin{figure}[htbp]
\subfigure[]{
\label{Crys}
\includegraphics[width=0.27\textwidth]{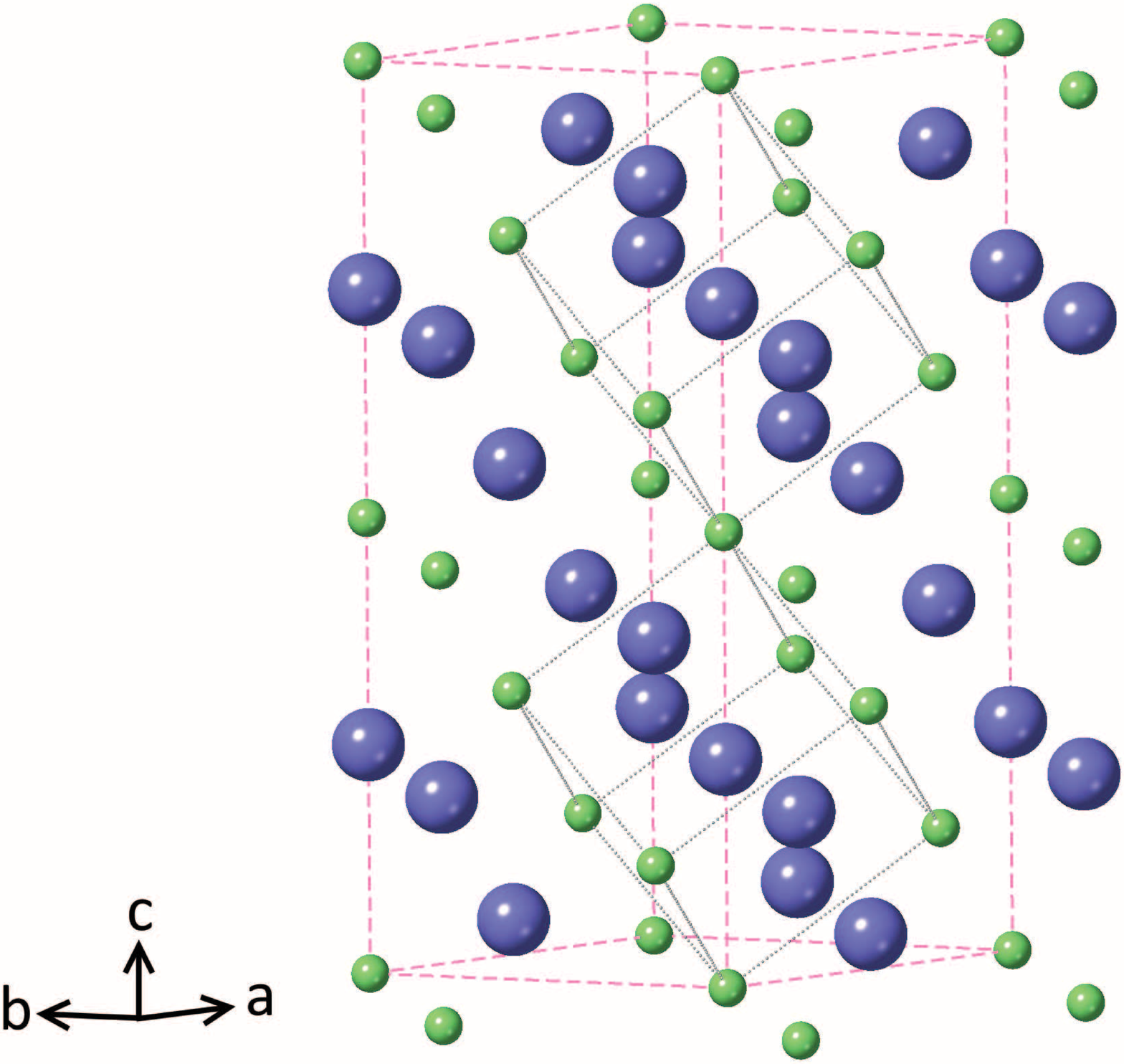}}
\subfigure[]{
\label{ResistivityM}
\includegraphics[width=0.19\textwidth]{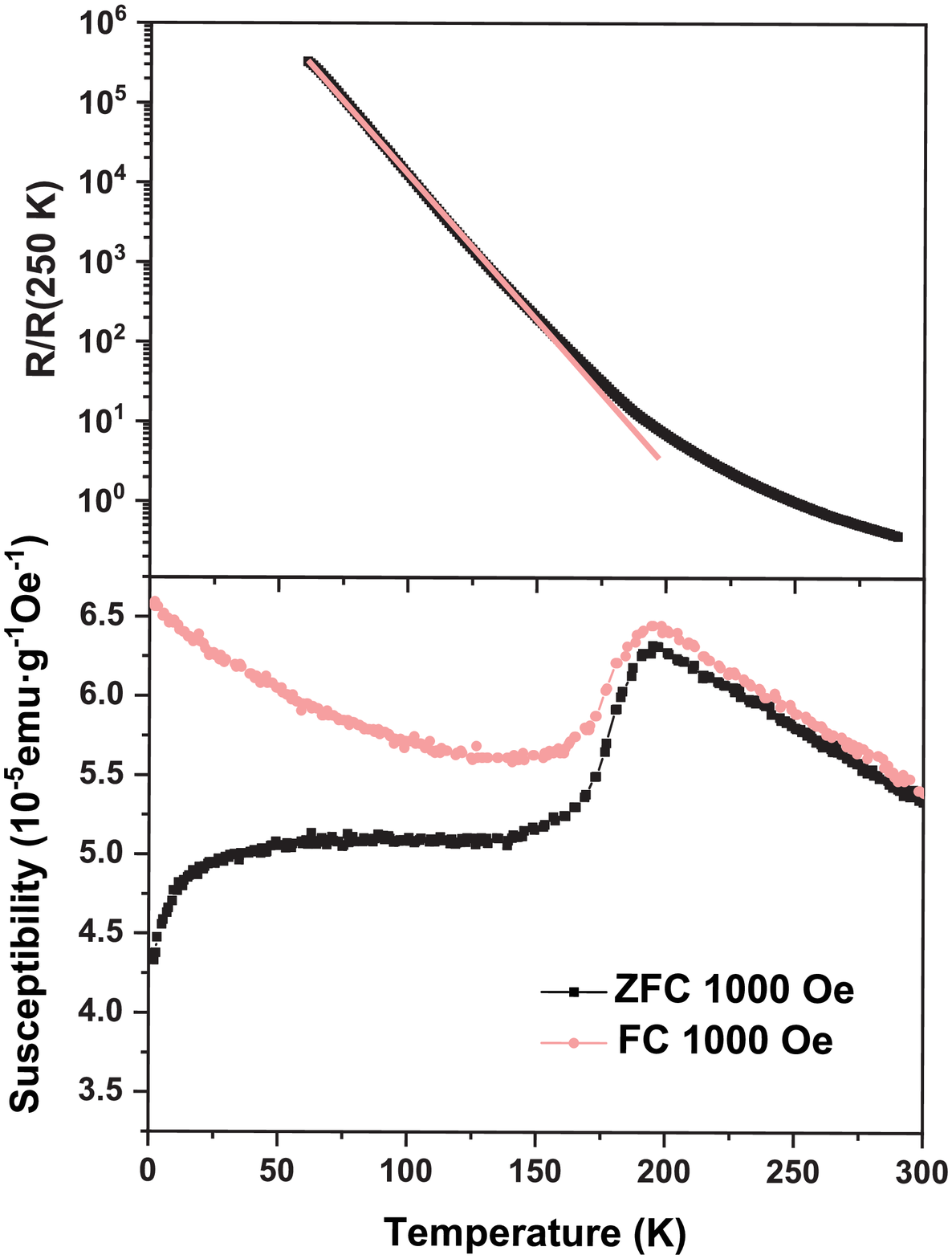}}
\caption{(Color online)(a) The crystal structure of La$_{1/3}$Sr$_{2/3}$FeO$_3$. The rhombohedral space group \hmn{R-3c} is shown in hexagonal setting (the unit cell in pink dot lines) and rhombohedral setting (the unit cell in grey dot lines). The hexagonal [001] is equivalent to the rhombohedral [111]. Purple balls denote La or Sr atoms and green balls denote Fe atoms. For clarity, oxygen atoms are not shown. (b) The temperature evolution of resistivity and magnetic susceptibility. The straight line drawn in the resistivity plotting is a guide to the eye.}
\label{CrysRM}
\end{figure}

\begin{figure*}[htbp]
\subfigure[]{
\label{magnetic_structure_P3_221}
\includegraphics[width=0.45\textwidth]{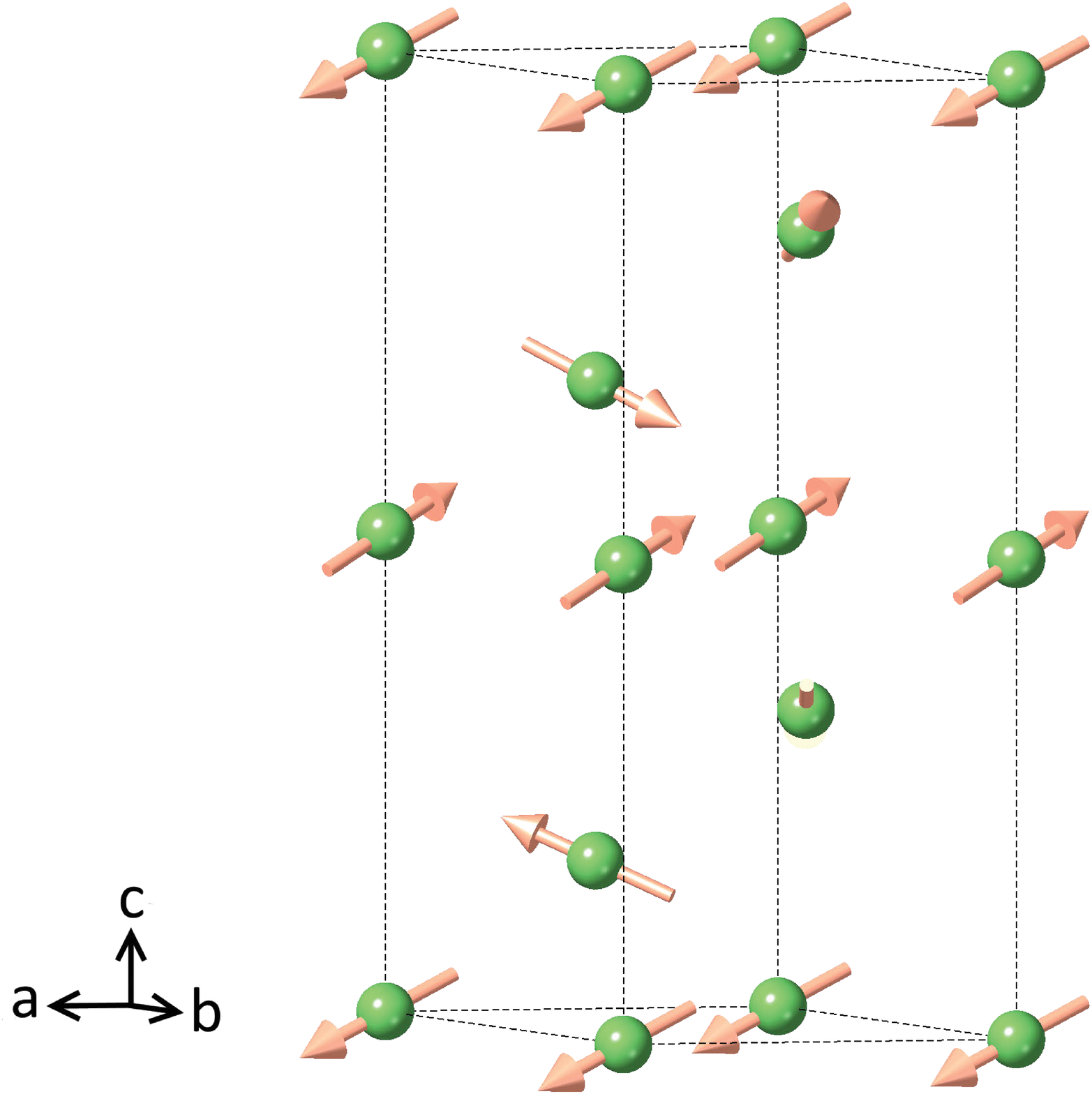}}
\subfigure[]{
\label{P3221_refinement}
\includegraphics[width=0.51\textwidth]{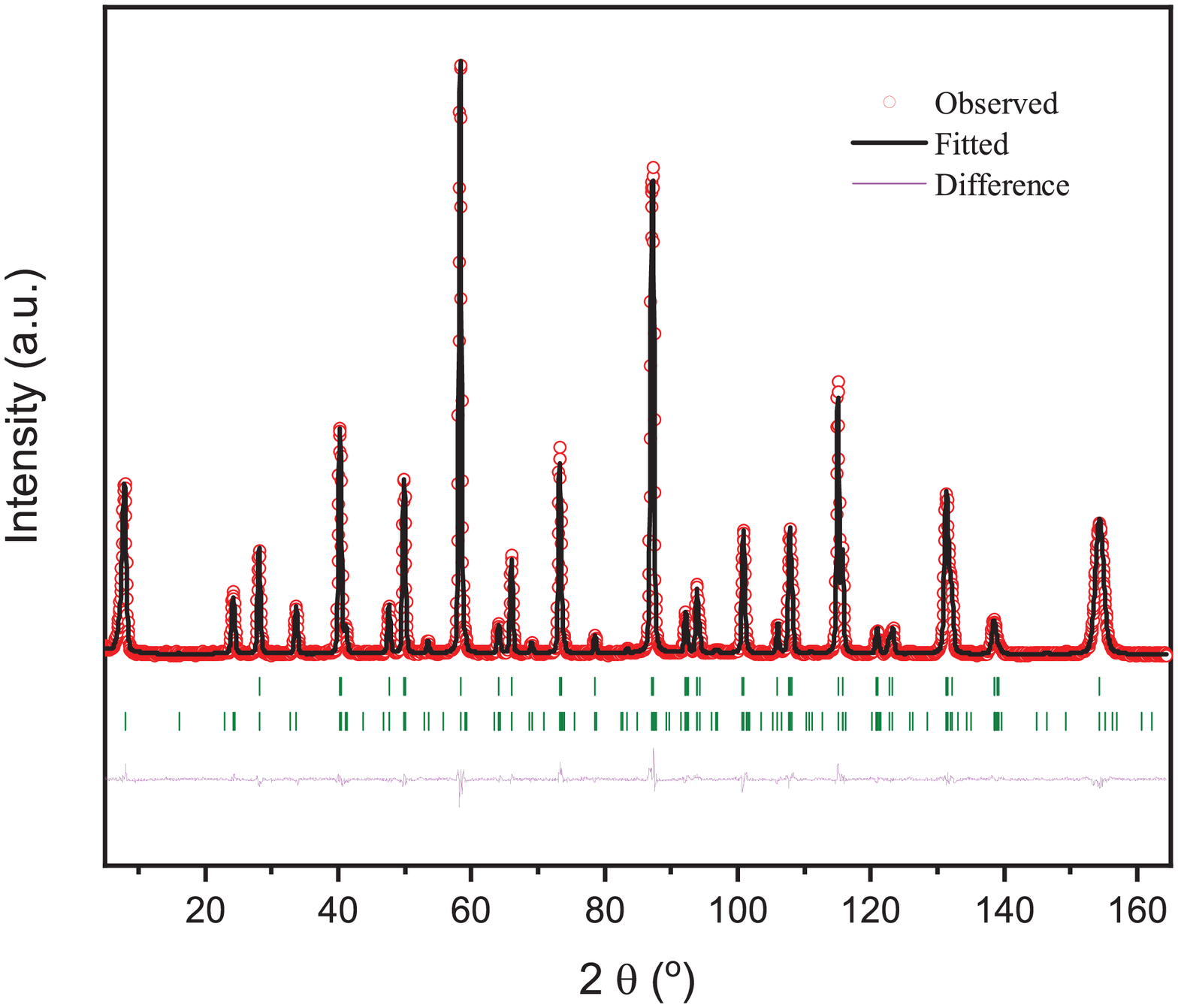}}
\caption{(Color online) (a) The helical magnetic structure of  La$_{1/3}$Sr$_{2/3}$FeO$_3$ at 2 K. (b) The Rietveld refinement of the neutron diffraction data of  La$_{1/3}$Sr$_{2/3}$FeO$_3$ collected on HRPT at 2 K ($\lambda$ = 1.89 {\AA}), based on the helical model. The top and bottom rows of ticks below the pattern are the Bragg peak positions for the nuclear and magnetic scattering, respectively. }
\label{helical}
\end{figure*}

The MI transition for R = La, Pr and Nd was explained by CD and CO. For R = La, the CO was found to occur by using M\"ossbauer spectroscopy \cite{Takano1981} and electron microscopy \cite{Li1997}. On the basis of the CO sequence ...-Fe$^{5+}$-Fe$^{3+}$-Fe$^{3+}$-..., the magnetic structure of this compound was reported to be \hmn{P-3m1} \cite{Battle1990} or \hmn{P1} \cite{Yang2003} from the neutron diffraction studies performed at 50 K and 15 K, respectively. The former seems not to be a correct solution since the presence of rotoinversion \hmn{-3} is incompatible with the claimed collinear magnetic structure, with the collinear moments in the \textit{ab}-plane in \hmn{R-3c} metric; and the latter might be a correct solution, but without any symmetry restrictions in space group \hmn{P1}. Moreover, the presence of Fe$^{5+}$ below $T_{\rm{MI}}$ is not consistent with the X-ray absorption data \cite{Blasco2008}, and resonant X-ray scattering measurements indicate that the CD is not significant \cite{Martin2009}. Furthermore, the R = Eu sample is reported to have a change of M\"ossbauer response aross the magnetic ordering transition similar to that of the R = La compound \cite{Huang2009}, which is surprising given the absence of MI transition. The change of the M\"ossbauer response for both compounds was then ascribed to the long-range magnetic ordering with two types of magnetic interactions\cite{Blasco2010}. Therefore the magnetic structure and the associated change of M\"ossbauer spectra are still not well understood yet.

In this paper we report neutron powder diffraction and M\"ossbauer spectroscopy studies in the temperature range 2-300 K for La$_{1/3}$Sr$_{2/3}$FeO$_3$. New models of magnetic structure are presented and their general implications and compatibility with the results of a local probe technique, $^{57}$Fe M\"ossbauer spectroscopy, are discussed.

\section{EXPERIMENTAl DETAILS}

The polycrystalline sample used in this study was prepared by solid state reactions. Stoichiometric amounts of dried La$_2$O$_3$, SrCO$_3$ and Fe$_2$O$_3$ were mixed thoroughly by hand in an agate mortar, placed in an alumina crucible and annealed at 1473 K for 40 h in a muffle furnace in the air. The obtained powder was then ground, pressed into a pellet and sintered at 1673 K for 40 h. The sintering was repeated once with intermediate grinding. To ensure the oxygen stoichiometry, the sample was further annealed under oxygen flow at 873 K for 72 h. Phase purity was checked by laboratory X-ray powder diffraction. The oxygen content was verified by thermogravimetric H$_2$ reduction analysis performed on Netzsch model STA 449C analyser. Resistivity and bulk magnetic properties were measured using a Quantum Design Physical Property Measurement System. The resistivity was measured on cooling and subsequently heating using the four-probe method. The magnetic susceptibility was measured using zero-field-cooled (ZFC) and field-cooled (FC) protocols.

The neutron diffraction data were collected at the Swiss Spallation Neutron Source (SINQ), Paul Scherrer Institute. Approximately 1 g of sample powder was loaded into a 6-mm-diameter vanadium can and the measurements were performed on the High-Resolution Powder Diffractometer for Thermal Neutrons (HRPT) \cite{Fischer2000} using $\lambda$ = 1.89 {\AA} and 1.15 {\AA} at 230 K and 2 K, and on the Cold Neutron Powder Diffractometer (DMC) using $\lambda$ = 2.46 {\AA} at a series of temperatures between 300 and 1.7 K. An absolute comparison on the 10$^{-3}$ level of crystal lattice parameters obtained from these two instruments is not possible, because of systematic uncertainties related to wavelength calibration and peak shape parameters. The neutron diffraction data were analyzed by Rieveld refinement using \emph{FULLPROF} suite \cite{Juan1993}, by using its internal tables of neutron scattering lengths and magnetic form factors. The symmetry analysis was done using the \textit{ISODISTORT} tool \cite{Campbell2006}, BasIreps option incorporated in \emph{FULLPROF} suite \cite{Juan1993} and software tools of the Bilbao crystallographic server \cite{Aroyo2011}.

\begin{table*}[htbp]
\caption{The matrices and basis vectors of the small irreducible representations for Fe in 6b position and \textbf{k}=(0,0,1), where $a = -\tfrac{1}{2}+\tfrac{\sqrt{3}}{2}i$, $b = -\tfrac{1}{2}-\tfrac{\sqrt{3}}{2}i$, $p = \tfrac{\sqrt{3}}{2}+\tfrac{1}{2}i$, $q = i$.
\label{Basireps table} }
\begin{ruledtabular}
\begin{tabular}{ccccccccc}
\multirow{2}{*}{}& \multirow{2}{*}{$\{1|000\}$} & \multirow{2}{*}{$\{3^{+}_{00z}|000\}$} & \multirow{2}{*} {$\{3^{-}_{00z}|000\}$} & \multirow{2}{*}{$\{m_{x-xz}|00\tfrac{1}{2}\}$} & \multirow{2}{*}{$\{m_{x2xz}|00\tfrac{1}{2}\}$} & \multirow{2}{*}{$\{m_{2xxz}|00\tfrac{1}{2}\}$}
 &  \multicolumn{2}{c}{Basis vector}\\
\cline{8-9}
&&&&&&&\textrm{Fe}$(0,0,0)$ & \textrm{Fe}$(0,0,1/2)$ \\
\colrule
$\Lambda$$_{1}$ & $1$ & $1$ & $1$ & $1$ & $1$ & $1$ & $(0,0,1)$ & $(0,0,-1)$\\
$\Lambda$$_{2}$ & $1$ & $1$ & $1$ & $-1$ & $-1$ & $-1$ & $(0,0,1)$ & $(0,0,1)$\\
$\Lambda$$_{3}$ & $\begin{bmatrix}  1 & 0 \\ 0 & 1 \end{bmatrix}$ &
               $\begin{bmatrix}  a & 0 \\ 0 & b \end{bmatrix}$ &
               $\begin{bmatrix}  b & 0 \\ 0 & a \end{bmatrix}$ &
               $\begin{bmatrix}  0 & 1 \\ 1 & 0 \end{bmatrix}$ &
               $\begin{bmatrix}  0 & b \\ a & 0 \end{bmatrix}$ &
               $\begin{bmatrix}  0 & a \\ b & 0 \end{bmatrix}$ &
                 \makecell{$(p^*,q^*,0)$ \\ $(0,0,0)$ \\ $(0,0,0)$ \\ $(p,q,0)$} &
                 \makecell{$(0,0,0)$ \\ $(q,p,0)$ \\ $(q^*,p^*,0)$ \\ $(0,0,0)$}\\
\end{tabular}
\end{ruledtabular}
\end{table*}

The $^{57}$Fe M\"ossbauer spectra were recorded in transmission geometry using a constant-acceleration spectrometer with a 25 mCi $^{57}$Co source in a Rh matrix. The velocity scale was calibrated with a metallic iron foil at room temperature. The data were analyzed with a least squares fitting program assuming Lorentzian peaks in the first-order approximation \cite{Caer}. Isomer shifts are given with respect to $\alpha$-Fe at room temperature.

\begin{figure}[bp]
\centering
\leavevmode
     \includegraphics[width=0.50\textwidth]{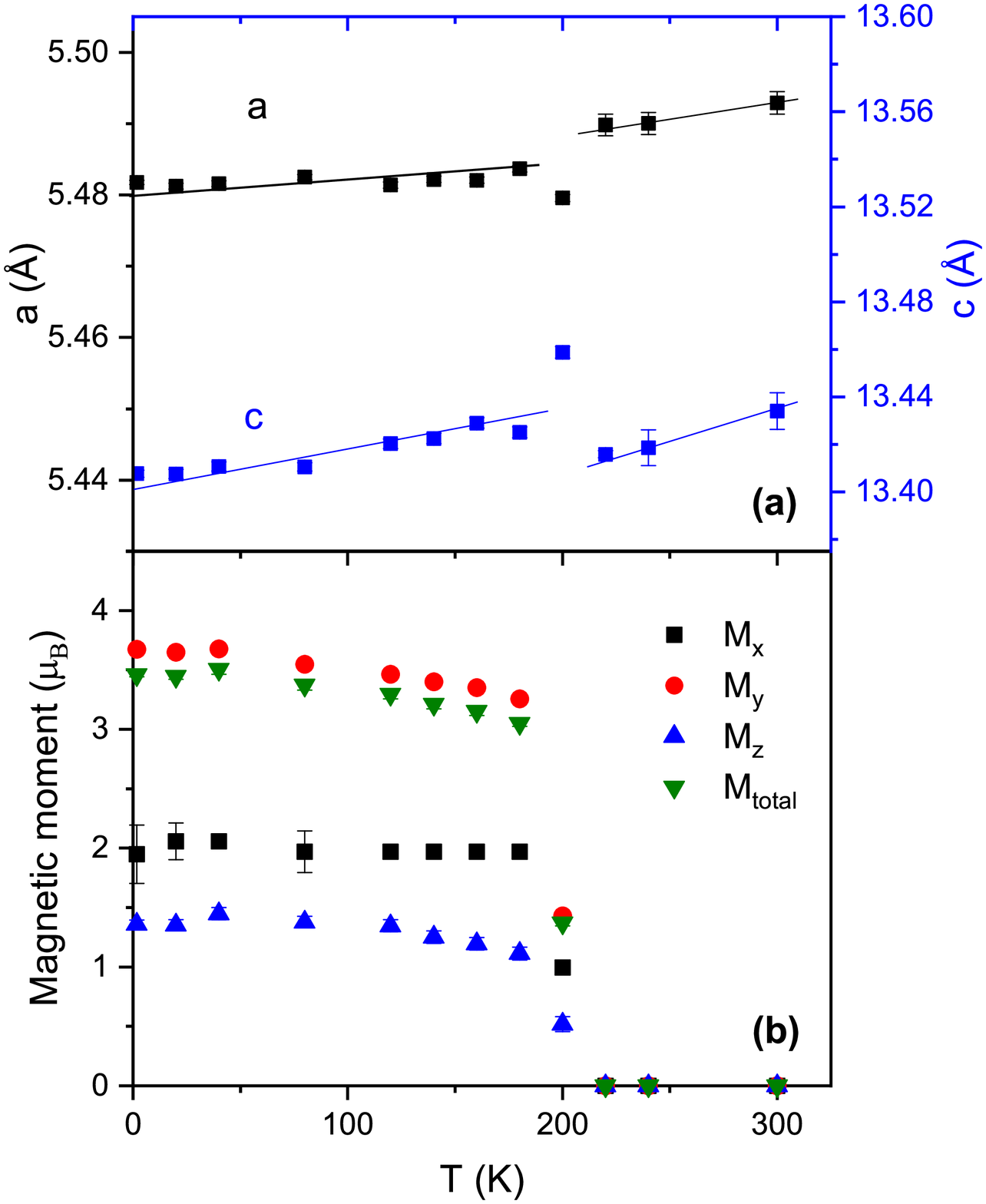}
 \caption{(Color online) The temperature evolution of (a) lattice parameters and (b) total magnetic moment and its components of Fe obtained from the Rietveld refinement of the neutron diffraction data of  La$_{1/3}$Sr$_{2/3}$FeO$_3$ collected on DMC, based on the helical model. The straight lines drawn in (a) are guides to the eye. If not visible, the error bars are smaller than the plotting symbols. See the text for details. }
 \label{Moment&lattice}
\end{figure}

\section{RESULTS AND DISCUSSION}

\subsection{Electric and magnetic properties}

La$_{1/3}$Sr$_{2/3}$FeO$_3$ crystallizes in \hmn{R-3c} space group at room temperature (see Fig. \ref{Crys}). In Fig. \ref{ResistivityM}, the temperature evolution of the resistivity, $R(T)/R(250~K)$, is presented.  A change of slope is visible at about 200 K, a temperature below which the material becomes more insulating. At this temperature a charge disproportionation is expected to take place. The transition observed here is less pronounced than that reported in \cite{Park1999, Zhao2000} on bulk samples, but it is very similar to that seen on thin films \cite{Okamoto2010}. This difference may arise from the oxygen stoichiometry of the sample. In our sample the oxygen stoichiometry is 3.02 $\pm$ 0.02.

An antiferromagnetic (AFM)-like transition is clearly observed at $T_{\rm{N}}$ $\sim$ 200 K in the DC magnetic susceptibility $\chi(T)$ measurement (see Fig. \ref{ResistivityM}), i.e., at the same temperature where a change of the slope in $R(T)/R(250~K)$ is observed. The data measured in ZFC mode diverge from that measured in FC mode below $T_{\rm{N}}$, suggesting that at low temperatures a spin-glass state or weak ferromagnetism develops.

\subsection{Magnetic and crystal structure}

\subsubsection{Symmetry analysis}

The neutron powder diffraction pattern shows the appearance of additional peaks below $\sim$ 200 K, which we interpret as magnetic scattering given the existence of a peak in the macroscopic magnetic susceptibility at this temperature (see Fig. \ref{ResistivityM}). The representation theory analysis has been performed in order to determine the magnetic structure at low temperatures, which is presented as follows.

\begin{table*} [htbp]
\caption{Crystal and magnetic structure parameters of La$_{1/3}$Sr$_{2/3}$FeO$_3$ in (a) parent paramagnetic space group \hmn{R-3c} (No. 167, hexagonal setting) at 300 K and in magnetically ordered state at 2 K in Shubnikov magnetic space group (b) \hmn{P3_{2}21} (No. 154.41 ) or (c) \hmn{C2/c} (No. 15.85). See text for more details.
\label{Structure parameters} }
\begin{ruledtabular}
\begin{tabular}{lccc}
& (a) \hmn{R-3c}, T = 300 K & (b) \hmn{P3_{2}21}, T = 2 K \footnote{Crystal structure parameters in the Shubnikov magnetic space group are derived from the parent group, according to the basis transformation from \hmn{R-3c} to \hmn{P3_{2}21} with a linear part (1,1,0), (-1,0,0), (0,0,1) and an origin shift (2/3,2/3, 1/12) and that from \hmn{R-3c} to \hmn{C2/c} with a linear part (1,-1,0), (1,1,0), (0,0,1) and an origin shift (0,0,0). The lattice parameters and the atomic displacement parameters B for \hmn{C2/c} are further refined.} & (c) \hmn{C2/c}, T = 2 K \footnotemark[1] \\
\colrule
 a ({\AA})& 5.48217(8)& 5.47754 & 9.49162(19)\\
 b ({\AA})&  &  & 5.47680(11)\\
 c ({\AA})& 13.40521(23) & 13.36215 & 13.36393(13)\\
 \textrm{\textbf{La1/Sr1}}  &  &  & \\
   \textrm{Wyckoff position} & 6\textit{a} & 3\textit{b} & 4\textit{e} \\
  \textrm{x, y, z} & 0, 0, 1/4 & 1/3, 0, 1/6 & 0, 0, 1/4 \\
  B ({\AA}$^2$)& 0.686(21) &  & 0.311(15)\\
 \textrm{\textbf{La2/Sr2}} &  &  & \\
    \textrm{Wyckoff position} &  & 3\textit{a} & 8\textit{f} \\
    \textrm{x, y, z} &  & 1/3, 0, 2/3 & 1/3, 0, -1/12 \\
  B ({\AA}$^2$)&  &  & 0.311(15) \\
  \textrm{\textbf{Fe1}} &  &  & \\
   \textrm{Wyckoff position} & 6\textit{b} & 6\textit{c} & 4\textit{a} \\
  \textrm{x, y, z} & 0, 0, 0 & 1/3, 0, 11/12 & 0, 0, 0 \\
  B ({\AA}$^2$) & 0.441(18) & 0.217 & 0.202(13)\\
  M$_x$($\mu_{\rm{B}}$), M$_y$($\mu_{\rm{B}}$), M$_z$($\mu_{\rm{B}}$) &  & 1.46(7), 3.67(2), 1.32(2) & 3.26(3), 0, 0 \\
  \textrm{\textbf{Fe2}}  &  &  & \\
  \textrm{Wyckoff position}&  &  & 8\textit{f} \\
  \textrm{x, y, z} &  &  & 1/3, 0, 2/3 \\
  B ({\AA}$^2$) &  &  & 0.202(13)\\
  M$_x$($\mu_{\rm{B}}$), M$_y$($\mu_{\rm{B}}$), M$_z$($\mu_{\rm{B}}$)  &  &  & -3.67(2), 0, 0 \\
  \textrm{\textbf{O1}} &  &  & \\
 \textrm{Wyckoff position}& 18\textit{e} & 6\textit{c} & 8\textit{f} \\
  x, y, z & 0.51812(18), 0, 1/4 & 1/3, 0.47410, 1/6 & 0.26295,0.26295, 1/4 \\
  B ({\AA}$^2$) & 1.057(19) &  & 0.572(12) \\
  \textrm{\textbf{O2}} &  &  & \\
  \textrm{Wyckoff position} &  & 3\textit{b} & 8\textit{f} \\
  \textrm{x, y, z} &  & 0.80743, 0, 1/6	& -0.07038, 0.26295, 7/12\\
  B ({\AA}$^2$) &  &  & 0.572(12)\\
  \textrm{\textbf{O3}} &  &  & \\
  \textrm{Wyckoff position} &  & 6\textit{c} & 8\textit{f}\\
  \textrm{x, y, z} &  & 1/3, 0.52590, 2/3 & 0.59628, 0.26295, -1/12 \\
  B ({\AA}$^2$)  &  &  & 0.572(12)\\
  \textrm{\textbf{O4}} &  &  & \\
  \textrm{Wyckoff position} &  & 3\textit{a} & 4\textit{e} \\
  \textrm{x, y, z} &  & 0.85923, 0, 2/3 & 0, 0.47410, 1/4 \\
  B ({\AA}$^2$)  &  &  & 0.572(12) \\
  \textrm{\textbf{O5}} &  &  & \\
  \textrm{Wyckoff position} &  &  & 8\textit{f} \\
  \textrm{x, y, z} &  &  & 1/3,  0.47410,  -1/12 \\
  B ({\AA}$^2$)  &  &  & 0.572(12) \\
\end{tabular}
\end{ruledtabular}
\end{table*}

\begin{figure*}[htp]
\subfigure[]{
\label{magnetic_structure_C2_c}
\includegraphics[width=0.45\textwidth]{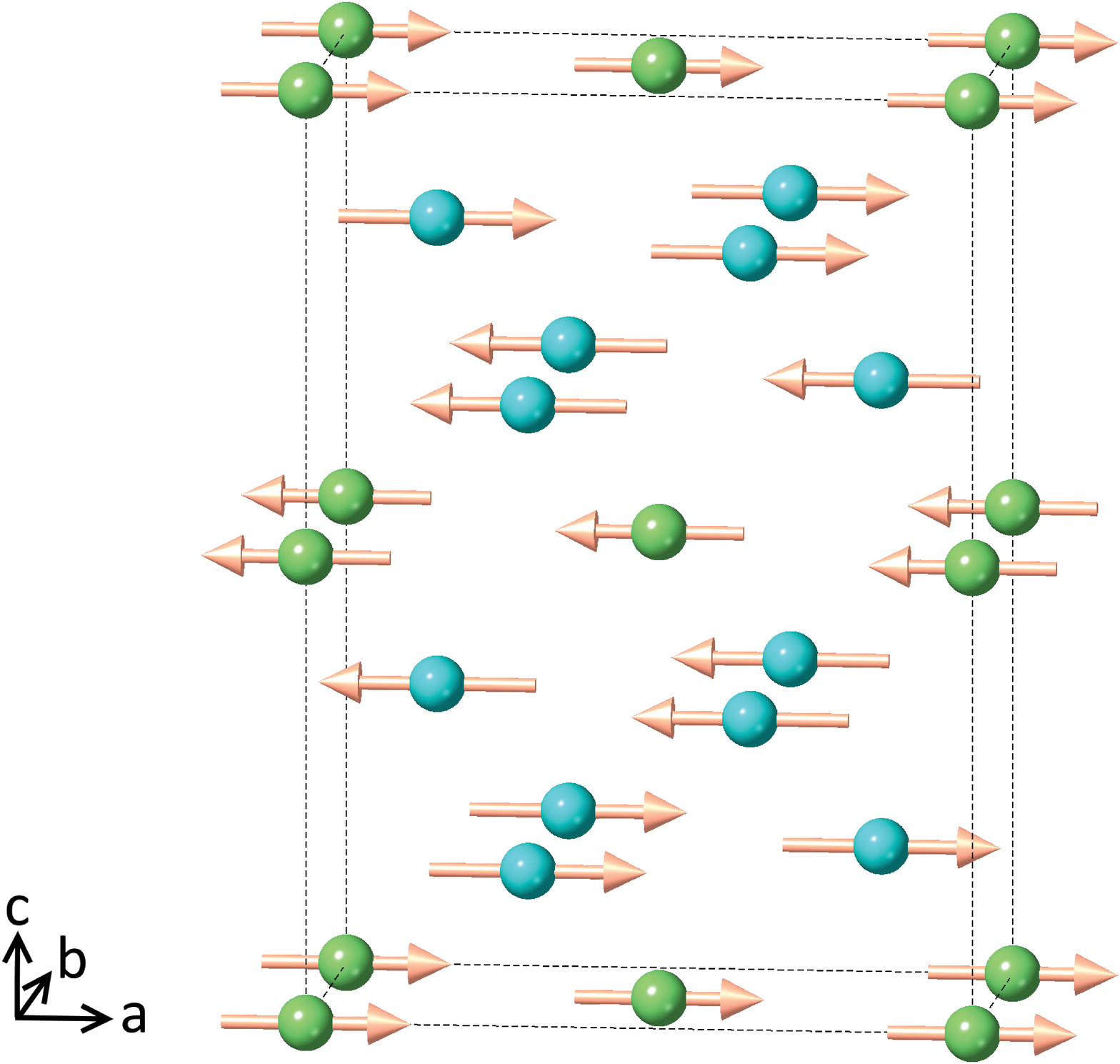}}
\subfigure[]{
\label{C2c_refinement}
\includegraphics[width=0.51\textwidth]{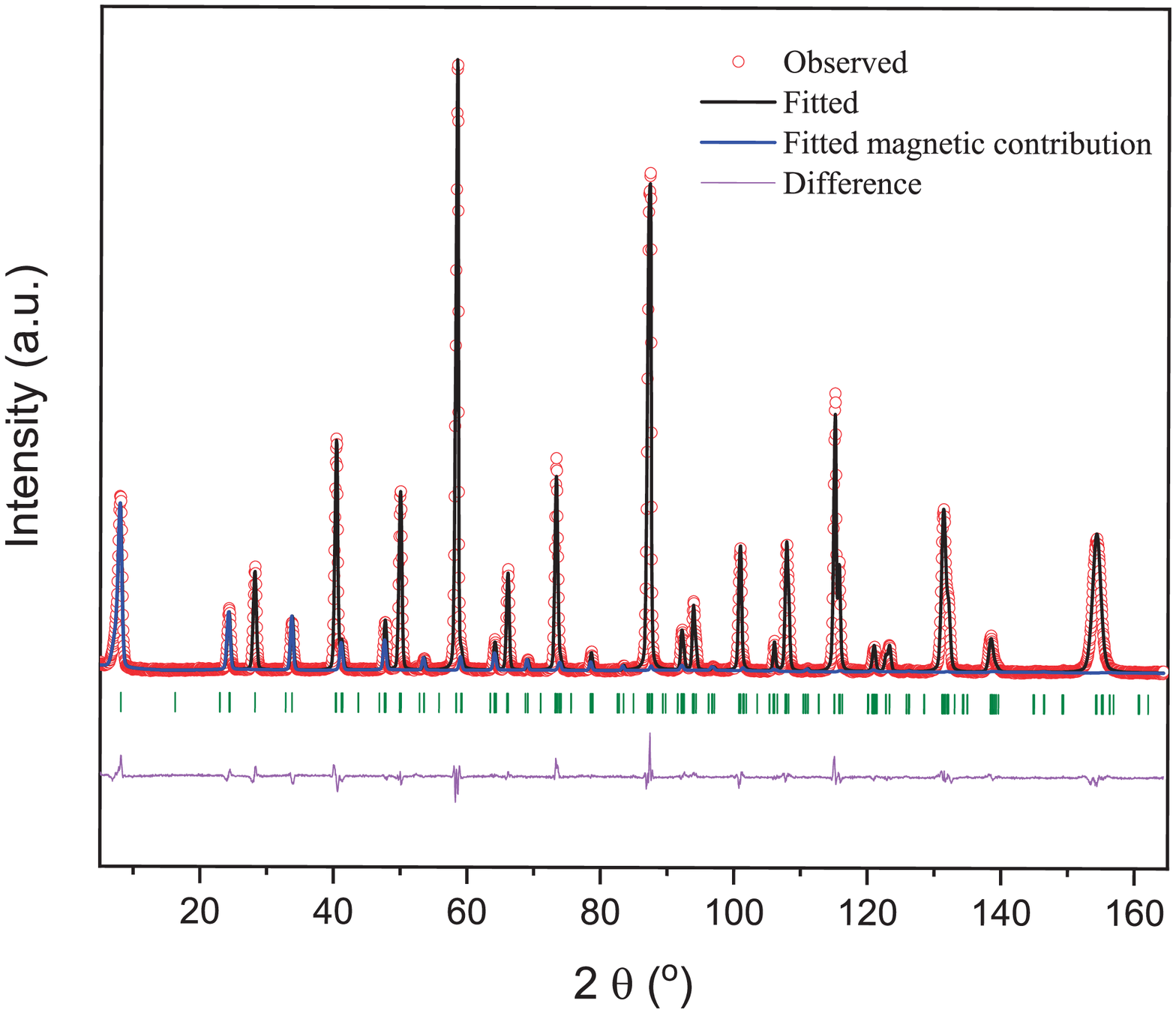}}
\caption{(Color online) (a) The collinear magnetic structure of  La$_{1/3}$Sr$_{2/3}$FeO$_3$ at 2 K. The green balls denote Fe$^{5+}$ and the blue balls denote Fe$^{3+}$. (b) The Rietveld refinement of the neutron diffraction data of  La$_{1/3}$Sr$_{2/3}$FeO$_3$ collected on HRPT at 2 K ($\lambda$ = 1.89 {\AA}), based on the collinear model. The ticks below the pattern are the Bragg peak positions for the nuclear and magnetic scattering. }
\label{collinear}
\end{figure*}

The magnetic order is considered to be characterized by a propagation vector \textbf{k}= (0,0,1) in \hmn{R-3c} metrics, as determined from the Le Bail fit. This is a model-free fit in which peak matching is tested with a certain propagation vector included as an additional phase. The propagation vector found here is the $\Lambda$ point of Brillouin zone, $\Lambda$ =(0,0,g), where g can have any value by symmetry, i.e. in general incommensurate. In this case it is considered to be locked to (0,0,1). It should be noted that this is not equivalent to $\Gamma$ point (0,0,0) because of the presence of R-centering translations. In primitive rhombohedral unit cell the propagation vector is \textbf{k$_p$}= (1/3, 1/3, 1/3). For \textbf{k}= $\Lambda$ in \hmn{R-3c} there are three possible small irreducible representations (irreps) of the \textbf{k}-vector group: $\Lambda$$_1$, $\Lambda$$_2$ and $\Lambda$$_3$, which are one-, one- and two-dimensional, respectively (we use nomenclature for irreps tabulated in \cite{Campbell2006}). For Fe in the 6\textit{b} (0,0,0) position the magnetic representation consists of $\Gamma_{mag} = 1 \Lambda_1 \oplus 1 \Lambda_2 \oplus 2 \Lambda_3$. These irreps and the corresponding basis vectors are listed in Table.\ref{Basireps table}. The $\Lambda$$_{1}$ and $\Lambda$$_{2}$ force the spin to be directed only along the \textit{c}-axis and have to be rejected, because of the presence of a strong (001)-magnetic peak in our experimental data. The solution is inevitably $\Lambda$$_{3}$. For this irrep all the basis vectors are in the \textit{ab}-plane. The irrep $\Lambda$$_3$ is 2-dimensional and enters two times in the magnetic representation. This, together with the fact that \textbf{k}-vector (0,0,g) is not equivalent to (0,0,-g) by symmetry, allows to reduce the symmetry even down to the space group \hmn{P1}. There are 14 different possible Shubnikov groups for a magnetic ordering according to irrep mLD3, found by \emph{ISODISTORT} software. Among them there are four maximal subgroups \hmn{P3_{2}21}, \hmn{P3_{2}2'1}, \hmn{C2/c} and \hmn{C2'/c'}. In the following we restrict the consideration to the maximal subgroups. There are two reasons for such restriction. Firstly, as will be shown below, the goodness of fit for some of them is as good as a Le Bail fit. Secondly, the latter two groups allow two Fe sites which could be compatible with the CO. It is worth to note that, the trigonal space groups \hmn{P3_{2}21} and \hmn{P3_{2}2'1} have their enantiomorphic pairs that should give equivalent description, namely \hmn{P3_{1}21} and \hmn{P3_{1}2'1}, respectively. The choice of space group between the pairs implies a particular domain choice. The enantiomorphic pair group corresponds to an equivalent structure related by the lost inversion center, and could have been equally used to describe the proposed magnetic structure.

\subsubsection{Helical model}

We first consider the most symmetric solution \hmn{P3_{2}21} for the irrep mLD3, which is generated by the order parameter (OP) direction mLD3 (0,0,a,0) \cite{Campbell2006}. It fits nicely to the neutron diffraction data ($\chi$$^2$ = 2.039, $R_{\rm{mag}}$ = 3.39$\%$). The magnetic R-factors are the same as that obtained for the Le Bail fit of the magnetic peaks where all peak intensities are treated independently. This implies that the above model cannot be improved any more. This model allows the presence of a secondary OP from an one-dimensional irrep mGM1+ in addition to the primary OP of mLD3. This results in an additional spin component along the \textit{c}-axis (see Ref. \cite{Gallego2016} for the general description of the symmetry concepts). This is a very good example of the case, where the combined irrep approach with the restriction coming from a particular magnetic space group consistent with the primary irrep gives a direct detection of the additional secondary component in the spin arrangement from the different irrep mGM1+. In the traditional approach that uses only irrep basis functions and is restricted in principle to a single irrep mLD3, this additional AFM canting would be impossible. The fit is considerably improved when the secondary mode mGM1+ is taken into account, as witnessed from the above goodness of fit indicators in comparison to $\chi$$^2$ = 5.170, $R_{\rm{mag}}$ = 10.50$\%$ when only a single irrep mLD3 is considered. The contribution from the secondary mode overlaps with that from the nuclear diffraction, but there is no correlation between them in the present case. Firstly, due to wide Q-range and only one free structure parameter (\textit{x}-position of oxygen atom) all nuclear contributions are practically fixed. Secondly, there are some peaks with significant contribution from \textit{c}-axis canting which are extinct for the nuclear phase due to \hmn{R-3c} symmetry, for instance the (011)-peak at 2$\theta$ = 24.4$^\circ$. We note that intensity of the above peak (and the other similar ones) is also zero for the main mLD3-component, providing convergence of the fit with secondary mode.

In this model, Fe cations are chirally arranged in the unit cell; all the moment directions are dictated by symmetry: the projection of the moments in the \textit{ab}-plane propagates helically along the \textit{c}-axis with \textbf{k}-vector $\Lambda$, and the moments projection on the \textit{c}-axis are antiferromagnetically stacked (see Fig. \ref{helical}). Only single Fe site is allowed by symmetry, with a magnetic moment of 3.46(2) $\mu_{\rm{B}}$ at 2 K (see Fig. \ref{Moment&lattice}). This model appears to exclude long-range CO or CD of Fe ions.

The magnetic moment of Fe obtained from the refinement of the DMC data evolves with temperature and shows a first-order like transition at $T_{\rm{N}}$. It shows no significant change below $T_{\rm{N}}$. The obtained value at 40 K and 20 K is respectively comparable to the averaged moment from Battle \textit{et al.}'s study ($\sim$3.31 $\mu_{\rm{B}}$ at 50 K) \cite{Battle1990} but much higher than that of Yang and coworkers ($\sim$ 2.43 $\mu_{\rm{B}}$ at 15 K)\cite{Yang2003}. The lattice parameters obtained from the refinement of the DMC data show a discontinuity at $T_{\rm{N}}$, which in this scenario may be ascribed to magnetostriction effects.

In the second trigonal group \hmn{P3_{2}2'1} the in-plane helical configuration is similar to that of \hmn{P3_{2}21}, but the secondary spin component is mGM2+ (ferromagnetic (FM) along \textit{c}-axis) and does not yield a convergent fit to the data.

\begin{figure}[bp]
\centering
\leavevmode
     \includegraphics[width=0.50\textwidth]{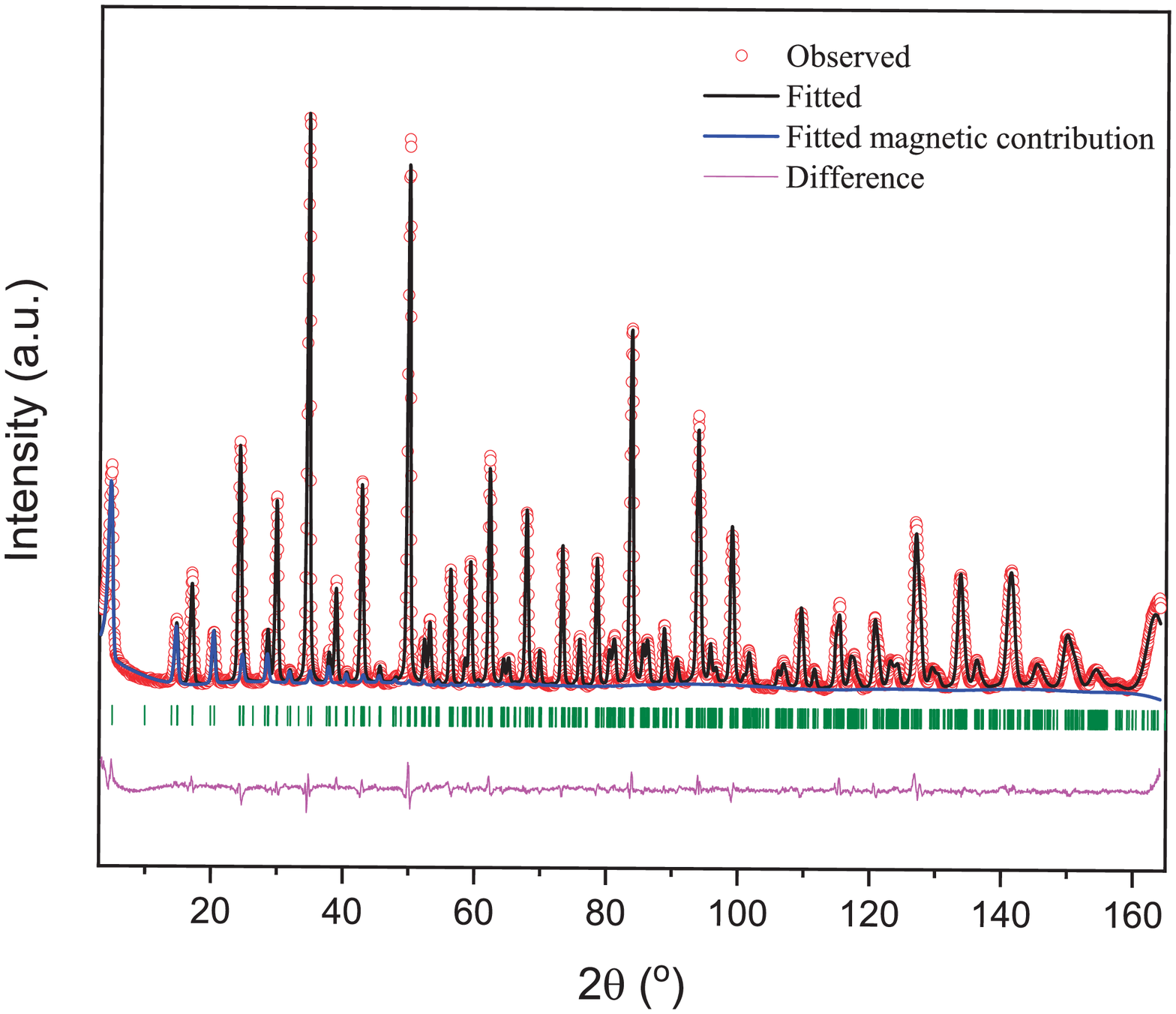}
 \caption{(Color online) The Rietveld refinement of the neutron diffraction data of  La$_{1/3}$Sr$_{2/3}$FeO$_3$ collected on HRPT at 2 K ( $\lambda$ = 1.15 {\AA}), based on the collinear \hmn{C2/c} model. The ticks below the pattern are the Bragg peak positions for the nuclear and magnetic scattering.}
 \label{C2c_1p15}
\end{figure}

\subsubsection{Collinear model}

Since CO was reported in the literature for this material \cite{Park1999}, we studied the less symmetric model that could be compatible with CO. The maximal symmetric subgroup would be \hmn{C2/c} and \hmn{C2'/c'}, generated by the OP direction mLD3 (0,0,a,a) and (a,-a,0,0) respectively, based on the propagation vector star (+$\Lambda$,-$\Lambda$). Both groups produce similar description of the experimental data: an amplitude modulation with two independent Fe moments. Both groups can produce the same spin configuration with however different moment direction: for \hmn{C2/c}, it is along \textit{a}-axis (shown in Fig. \ref{collinear}) while for \hmn{C2'/c'} it is along \textit{b}-axis (not shown). The spins of Fe ions are aligned collinearly. The couplings are FM between the ions of different charge and AFM between those of the same charge. In both cases this spin configuration is generated by mLD3 and mGM3+ irreps, the latter being a secondary OP. The contribution of the $\Gamma$ point is important, because it not only improves the fitting of the magnetic peaks, but also allows the proposed CO sequence. The magnetic configuration for other spin component are similar but not the same in both groups and releasing them does not give a convergent fit, as we explain below in this section. In the following we show only the results for the case of \hmn{C2/c}. The position of Fe splits up from 6\textit{b} in \hmn{R-3c} into 4\textit{a} and 8\textit{f} in \hmn{C2/c} (see Table \ref{Structure parameters}). When only mLD3 is considered, the fitting of the magnetic peaks of the neutron diffraction pattern at 2 K is poor ($\chi$$^2$ = 3.093, $R_{\rm{mag}}$ = 9.39$\%$). It gives an AFM spin configuration similar to that shown in Fig. \ref{collinear}, however the CO sequence as suggested from the relative moment size is ...-Fe$^{3+}$-Fe$^{5+}$-Fe$^{5+}$-..., which is not consistent with the previous studies \cite{Battle1990,Yang2003,Park1999}. The mGM3+ mode may give moments along the \textit{a}-axis. When it is taken into account, the magnetic peaks can be fitted well (see Fig. \ref{collinear}). The fitting yields $\chi$$^2$ = 4.140, $R_{\rm{mag}}$ = 4.26$\%$, slightly worse than that for the helical model. In this model, there are two Fe sites dictated by the space group symmetry with the CO sequence ...-Fe$^{5+}$-Fe$^{3+}$-Fe$^{3+}$-..., where Fe$^{3+}$ and Fe$^{5+}$ correspond to 8\textit{f} and 4\textit{a} positions in \hmn{C2/c} symmetry, respectively. The refined magnetic moment of the nominal Fe$^{5+}$ and Fe$^{3+}$ is 3.26(3) $\mu_{\rm{B}}$ and 3.67(2)$\mu_{\rm{B}}$, respectively. The average magnetic moment is 3.53 $\mu_{\rm{B}}$, comparable to the value of the single moment obtained from the helical model.

We also studied in more details the possibility to have other components for the Fe spins, and in particularly the \textit{c}-canting similar to that in the helical model. The components of magnetic moment are possible along \textit{a}, \textit{b} and \textit{c}-axes in \hmn{C2/c}. The AFM configuration for Fe$^{5+}$ and Fe$^{3+}$ spins shown in Fig. \ref{collinear} is possible only along \textit{a}-axis. The component along \textit{b}-axis is FM, and the component along \textit{c} is AFM. In \hmn{C2/c} group similar to \hmn{P3_{2}21} it is possible to have secondary symmetry modes from $\Gamma$ point mGM3+ (as explained above in this section) and/or mGM1+. The irrep mGM1+ gives the same AFM structure of \textit{c}-component as for the helical model for both Fe1 and Fe2 sites together. We attempted to fit in this model but there was no convergence. The convergence could not be reached either when the component along \textit{b}-axis was further released for refinement.

\begin{table*}[htbp]
\caption{The hyperfine parameters of La$_{1/3}$Sr$_{2/3}$FeO$_3$ at 300 K, 200 K and 4 K: line width $\Gamma$, isomer shift $\delta$, apparent quadrupole splitting 2$\epsilon$, and hyperfine field $H$. }
\label{Hyperfine parameters}
\begin{ruledtabular}
\begin{tabular}{cccccc}
 & \textrm{Proportion(\%)}& $\Gamma$ (mm/s) & $\delta$ (mm/s) &  2$\epsilon$ (mm/s) & $H$ (T)\\
\colrule
300 K & 100 & 0.37(2) & 0.13(2) & - & -\\
200 K & 100 & 0.35(2) & 0.20(2) & - & -\\
\makecell{4 K, site A\\4 K, site B} & \makecell{64.9\\33.3} & \makecell{0.36(2)\\0.31(2)} & \makecell{0.38(2)\\-0.02(2)} & \makecell{0.00(2)\\0.00(2)} & \makecell{46.4(2)\\26.5(2)}\\
\end{tabular}
\end{ruledtabular}
\end{table*}

\begin{figure}[bp]
\centering
\leavevmode
     \includegraphics[width=0.50\textwidth]{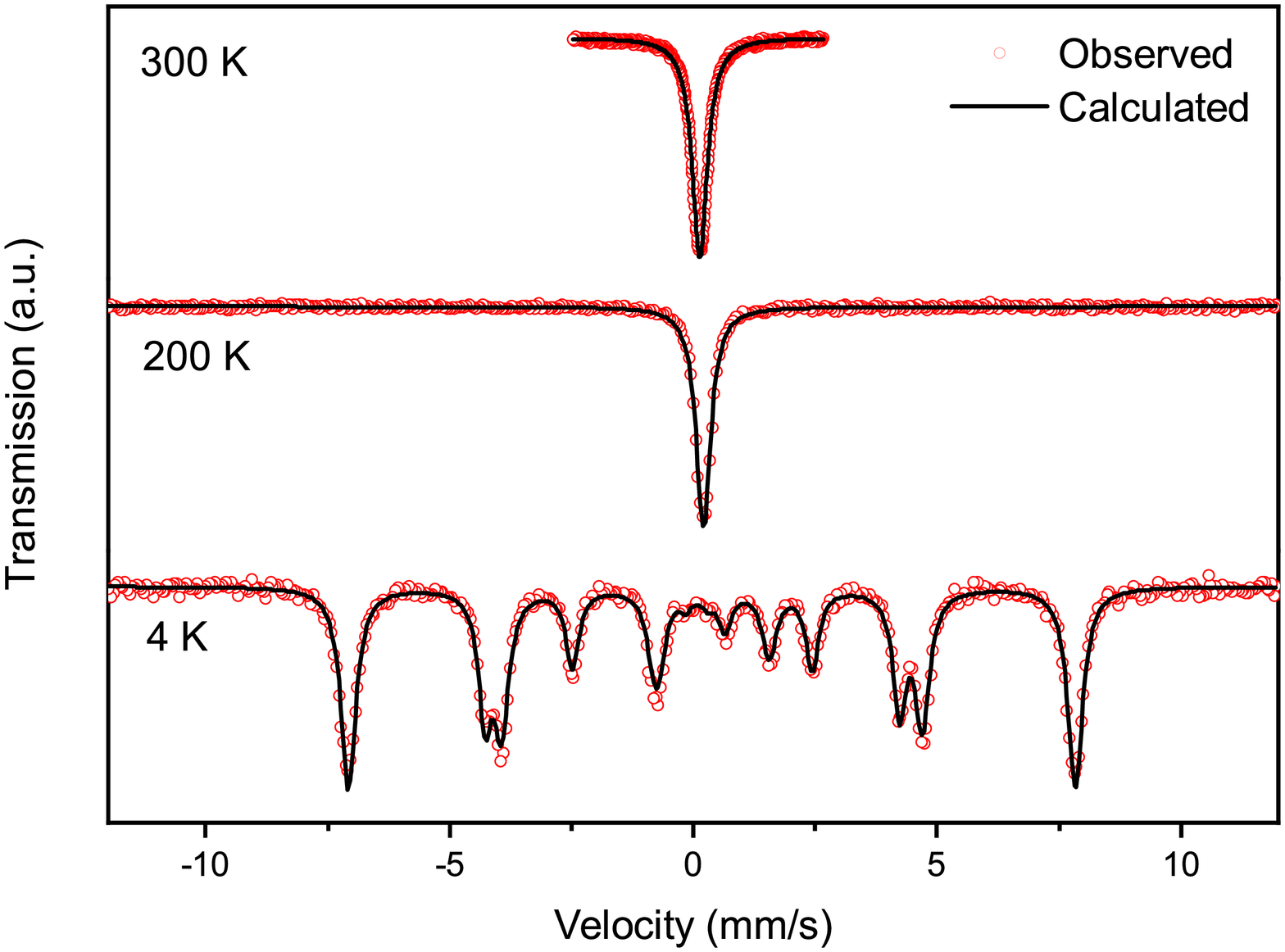}
 \caption{(Color online) The M\"ossbauer spectra of La$_{1/3}$Sr$_{2/3}$FeO$_3$ measured at 300 K, 200 K and 4 K.}
 \label{Mossbauer}
\end{figure}

The magnetic moment on the Fe$^{3+}$ site is larger than that on the Fe$^{5+}$ site, but the difference in amplitude is much smaller than expected for 3+ and 5+ valences. Thus this model does not support an ideal CO, but does not exclude a partial CD. In the present case of propagation vector \textbf{k$_p$}= (1/3, 1/3, 1/3), the deviation of the crystal structure from the paramagnetic \hmn{R-3c} symmetry should result in additional satellite reflections appearing at the same positions as the magnetic satellites. This makes the separation of nuclear and magnetic contributions more difficult. However the intensities of the structural satellites will not be suppressed by the magnetic form factor at large values of momentum transfer Q. A neutron diffraction pattern measured at $\lambda$ = 1.15 {\AA} allows us to go up to Q$_{max}$ = 11 {\AA}$^{-1}$. A detailed inspection of the measured pattern did not reveal the presence of any separate isolated diffraction peaks allowed in \hmn{C2/c} space group at high Qs (see Fig. \ref{C2c_1p15} ). We tentatively tried to release the atomic positions in \hmn{C2/c} model from the ideal average positions given by \hmn{R-3c} paramagnetic group (see Table \ref{Structure parameters}), but we were not able to obtain a convergent fit.

\subsection{Hyperfine structure}

The $^{57}$Fe M\"ossbauer spectra recorded at 300 K, 200 K and 4 K are shown in Fig. \ref{Mossbauer}. The spectrum at 300 K comprises a single broad line. Since the deviation from local cubic symmetry at the Fe site is very weak, it has been fitted using a singlet. The fitted hyperfine parameters are given in Table \ref{Hyperfine parameters}. Note that fittings using a doublet yield an isomer shift ($\delta$ $\sim$ 0.13 mm/s) identical to that when using a singlet, and a very small quadrupolar interaction ($\Delta$ $\sim$ 0.12 mm/s) with a slightly reduced linewidth ($\Gamma$ $\sim$ 0.32 mm/s). The spectrum recorded at 200 K, just above $T_{\rm{MI}}$, can be fitted in the same manner as the room temperature one. The $\delta$ value above $T_{\rm{MI}}$ lies in between that expected for Fe$^{3+}$ and Fe$^{4+}$. It thus agrees with the formal charge Fe$^{3.66+}$ deduced from the chemical formula.

The spectrum recorded at 4 K comprise two sextets of unequal intensity that correspond to two M\"ossbauer sites, A and B. The fitted hyperfine parameters are given in Table \ref{Hyperfine parameters}. They are in good agreement with those of previous M\"ossbauer studies \cite{Takano1981,Battle1988,Gallagher1967,Chernyshev2014}. In addition, minor non-magnetic contributions in the central part of the velocity scale were also taken into account, however they represent less than 2\% of the resonant area.

The two sextets have different isomer shifts and hyperfine fields. They thus correspond to different Fe charge states. In contrast with conclusions drawn from previous M\"ossbauer studies \cite{Takano1981,Battle1988}, the less intense sextet (Fe B site, $\sim$ 34\%) does not correspond to the rare Fe$^{5+}$ charge state; because its isomer shift ($\sim$ -0.02 mm/s) is not negative enough. The $\delta$ and \textit{H} values are however lower than those of Fe$^{4+}$ in SrFeO$_3$ ($\delta$ $\sim$ 0.146; $H$ $\sim$ 33.1 T)\cite{Gallagher1964}, suggesting that site B corresponds to a non-integer charge state intermediate between Fe$^{4+}$ and Fe$^{5+}$. Similarly, the hyperfine parameters of site A ($\delta$ $\sim$ 0.38 ; $H$ $\sim$ 46.4 T), whose spectral weight is twice that of the Fe B site, do not correspond to that of ‘pure’ Fe$^{3+}$ as in $\alpha$-Fe$_2$O$_3$ ($\delta$ $\sim$ 0.48 ; $H$ $\sim$ 54 T)\cite{Yoshida2013}. This suggests that the Fe A site has also a non-integer charge state, slightly higher than trivalent. We hence conclude that the charge difference below $T_{\rm{MI}}$ is rather limited, involving two iron sites with non-integer charge states Fe$^{(3.66-\zeta)+}$ and Fe$^{(3.66+2\zeta)+}$ for Fe A and Fe B, respectively. Although the Fe charge states cannot be determined precisely, we estimate 0.2 \textless $\zeta$ \textless 0.5. This agrees with the conclusion of the M\"ossbauer study in ref. \cite{Chernyshev2014}. From electronic spectroscopy data, Herrero-Martin \textit{et al.}\cite{Martin2009} also concluded to a modest charge segregation. However, their conclusions included that the higher charge state has twice the spectral weight of the lower one, which is not consistent with the present and past \cite{Takano1981,Battle1988,Gallagher1967,Chernyshev2014} M\"ossbauer data.

\subsection{Discussion}

The collinear model seems to be consistent with the present and previous results of M\"ossbauer spectroscopy \cite{Takano1981,Chernyshev2014} and the previous electron diffraction study \cite{Li1997}. The M\"ossbauer data can be simply analyzed by considering that the Fe$^{3.66+}$ disproportionates below $T_{\rm{MI}}$ into two Fe sites: Fe$^{(3.66-\zeta)+}$ and Fe$^{(3.66+2\zeta)+}$, in the ratio 2:1. This agrees with the collinear model with the two Fe sites in 8\textit{f} and 4\textit{a} positions corresponding to the M\"ossbauer sites A and site B, respectively. However, the magnetic moments obtained from the refinement of the neutron diffraction data are not fully consistent with the fitted hyperfine field values. The hyperfine field is built up from several contributions: the Fermi contact field (valence and core), the dipolar field and the orbital field \cite{Novak2010}. Although only the core contribution to the Fermi contact field scales with the magnetic moment, the hyperfine field to magnetic moment ratio generally lies in the 10-15 T/$\mu_{\rm{B}}$ range for Fe. Hence, the Fe moment deduced from the M\"ossbauer hyperfine field at sites A (8\textit{f}) and B (4\textit{a}) lies in the range between $\sim$ 3.1-4.6 $\mu_{\rm{B}}$ and $\sim$ 1.8-2.7 $\mu_{\rm{B}}$, respectively. The refined magnetic moment at the 4\textit{a} position is significantly higher (3.26 $\mu_{\rm{B}}$), which would imply a conversion factor as low as $\sim$ 8 T/$\mu_{\rm{B}}$.

The helical model appears inconsistent with the above results, however, it may not be fully ruled out. One possibility could be that electronic relaxations occur between two charge states at all temperatures. Relaxations are fast above the transition hence a single state is observed; while below $T_{\rm{MI}}$, they slow down and become slower than the M\"ossbauer probing time (10$^{-7}$ s) thus the two charge states are resolved. In this way, the charge separation below $T_{\rm{MI}}$ might only be apparent and only a single Fe site can be observed from neutron diffraction. The mean hyperfine field ($\sim$ 39.6 T) and the refined iron moment in the helical model (3.46 $\mu_{\rm{B}}$) yields a conversion factor of $\sim$ 11.4 T/$\mu_{\rm{B}}$ which lies in the commonly valid range of 10-15 T/$\mu_{\rm{B}}$. The other possibility could be that at low temperatures Fe cations could have two different valences locally, hence this can be probed by M\"ossbauer spectroscopy and electron diffraction, however the CO may not be long-ranged. Moreover the electrons are partially itinerant below $T_{\rm{MI}}$, which implies that a short-range CO is more likely. It is also worthwhile to add that a helical model is considered to be more energetically favorable than a collinear AFM state \cite{Shraiman1989,Luscher2007}. A spiral structure was proposed for the spin-glass state of La$_{2-x}$Sr$_x$CuO$_4$ \cite{Luscher2007}. In La$_{1/3}$Sr$_{2/3}$FeO$_3$, such a spin-glass ground state is also possible (see Fig. \ref{ResistivityM}).

Next we would like to point out some implications of the one-Fe helical model. It suggests that the first-order like MI transition is driven purely by magnetic ordering. This is in qualitative agreement with a recent experimental observation \cite{Devlin2014}: a negative magnetoresistance and a sign reversal of the Hall effect below $T_{\rm{MI}}$ is reported for R = La, and the exotic low-temperature transport properties are ascribed to a consequence of the unusually long-range periodicity of the AFM ordering.

\section{SUMMARY AND CONCLUSIONS}

The low-temperature magnetic structure of La$_{1/3}$Sr$_{2/3}$FeO$_3$ have been revisited and studied by neutron powder diffraction and a complementary M\"ossbauer spectroscopy. Based on the symmetry analysis, two crystallographic magnetic models, namely a chiral helical maximal symmetry \hmn{P3_{2}21} and a collinear \hmn{C2/c} or \hmn{C2'/c'} model are proposed. We found both models fit equally well with the neutron diffraction pattern at 2 K. The less symmetric \hmn{C2/c} or \hmn{C2'/c'} model allows charge ordering of Fe ions but our experimental data do not show any evidence of the expected structural distortion. The M\"ossbauer spectroscopy results appear to support the collinear model but cannot fully rule out the helical one. The latter model suggests that the metal-insulator transition is of magnetic origin. Polarised neutron diffraction on single crystals is needed to verify the validity of either of the models.

\begin{acknowledgments}

F.L and R.Y acknowledge the financial support from the SNSF (Schweizerischer Nationalfonds zur F\"orderung der Wissenschaftlichen Forschung) (Grant No.200021\_157009). R.Y also acknowledges the financial support from Horizon 2020 (Grant No.654000). We thank Prof M. Kenzelmann for fruitful discussions. The work was partially performed at the Swiss Spallation Neutron Source SINQ (PSI).

\end{acknowledgments}

\bibliography{LaSrFeO3}

\end{document}